\documentclass[12pt]{article}
\usepackage{amssymb,amsmath,mathrsfs}

\newcommand{\like}{\mathscr{L}}
\newcommand{\project}[1]{\textsl{#1}}
\newcommand{\LCDM}{$\Lambda$CDM}
\newcommand{\latin}[1]{\textit{#1}}
\newcommand{\etal}{\latin{et~al.}}

\sloppypar

\begin{document}
\section*{Is cosmology just a plausibility argument?\footnote{%
  A contribution to the meeting
  \textit{Exploring the High Energy Universe}
  in honor of Roger Blandford.
  Published on the arXiv only.
  Copyright 2009 David W. Hogg.
  You may copy and distribute this document
  provided you make no changes to it whatsoever.}}
\noindent
  David W. Hogg \\
  \textsl{Center for Cosmology and Particle Physics} \\
  \textsl{Department of Physics, New York University} \\
  \texttt{david.hogg@nyu.edu}

\paragraph{Abstract:}
I review the basis and limitations of plausible inference in
cosmology, in particular the limitation that it can only provide
fundamentally \emph{true} inferences when the hypotheses under
consideration form a set that is \emph{exhaustive}.  They never do;
this recommends abandoning \emph{realism}.  Despite this, we can adopt
a scientifically correct pragmatism and understand aspects of the
cosmological model with enormous confidence.  I illustrate these
points with discussion of one certainty---expansion---and two current
controversies---the existence of large extra dimensions and the
possibility that the matter distribution forms a fractal on large
scales.  I argue that the existence of large extra dimensions is
certainly plausible, but a fractal universe is untenable.

\section{Radicals, plausibility, and \textsl{The Beagle}}

In my time in the nineteen-nineties working as a student of Roger
Blandford, he asked several things of me that have rung in my head.
At that time the matter density of the Universe could have been
$\Omega=0.1$ or $\Omega=1.0$, the Hubble Constant could have been 50
or 100 (in the usual units), hot dark matter was as good as cold, the
initial conditions could have been adiabatic or isocurvature or worse,
and only a fringe believed there might be cosmic acceleration.
Fluctuations were discovered in the cosmic microwave background, the
ten-meter telescopes were coming on-line, the highest known galaxy
redshifts jumped from 0.4 to 3, and gamma-ray bursts became
confidently cosmological.  Among other things, Blandford and I hoped
to constrain some of the fundamental properties of the Universe with
\project{Keck} and \project{Palomar} observations of hundreds of
galaxies in the Hubble Deep Field and various other ``selected
areas''.\footnote{We failed, although I, for one, had a great time.}
Here are three things that Blandford asked me:

\textsl{(1)}~``Where are the young radicals?''  In this period, the
notable objectors to cosmological orthodoxy included Arp, Burbidge,
Hoyle, and Segal, proponents of quasars as galaxy ejecta, steady-state
models, periodic redshifts, chronometric cosmology, and a range of
other relatively untenable ideas.  Even among the main-stream, those
over the age of 60 were much more likely to fairly discuss a crazy
idea about the Universe than anyone under the age of 40.  When the
youth---those who are the engines of experimental research, data
analysis, and new projects---don't care about outside chances, the
outside chances never get properly tested, ruled out, or investigated
for the seeds of new and more promising ideas.  That's not healthy.

\textsl{(2)}~``Why don't you gather up all the phenomenology, pack
your carpet bag, and sail around the Cape in \textsl{The Beagle}?''
Phenomenology was pouring in from telescopes across the spectrum and
in such detail that absolutely \emph{no} theoretical model could be
consistent with even a small fraction of it.  This continues today: No
theoretical model of galaxies simultaneously explains all the rich
phenomenology in the scaling relations, mass functions, star-formation
rates, chemical abundances, morphologies, and clustering.  Why are we
taking so much more data when we can't even make good progress on what
we have?  Stop gathering and publishing incremental snippets of
confusing phenomena (was the implication) and start trying to
understand how it all fits together; this will require a long period
of concentration!  When you get back from the trip, write \emph{The
  Origin of Galaxies} and begin a period of maturity for observational
cosmology.

\textsl{(3)}~``Isn't that just a \emph{plausibility argument}?''  Is
the galaxy auto-correlation function (for example) a power law because
the underlying dark-matter structure forms a power-law, or because
auxilliary aspects of galaxy formation conspire to make a
non-power-law dark-matter auto-correlation function into a power-law
galaxy auto-correlation function?  Here are two explanations of the
same phenomenon.  How to decide between them, when (at the time) there
is (was) no means to distinguish them directly with an
observation?\footnote{This is just an example; of course now we know
  the answer from weak lensing studies (for example, Sheldon
  \etal\ 2009).}  Any argument about this was purely in the realm of
plausibility; both sides agree on the phenomena, so the differences of
scientific conclusion are just differences in what seems plausible.
Now, is \emph{everything} in astronomy---where we can't do controlled
or repeated experiments, and have (almost) exclusively electromagnetic
channels for observation---a plausibility argument?  Can we
\emph{know} anything?

I will argue in what follows that these three questions are
related. The short answer to the last one is ``yes''; observational
science is a science of plausibility, and plausibility arguments are
an unavoidable part of doing business.  Nonetheless, as you know---and
as I will try to complexify---we know many things in cosmology with
great certainty.

\section{Bayesian science}

There is an unfortunate set of battles going on in physics about
statistics---``frequentism'' against ``Bayesianism''---that are
distorting the meanings of both terms.  Although if you read my recent
papers you will know who I support in this war, I do \emph{not} want
to discuss this question here.  I want to take back the word
``Bayesian''; I want to use it to describe a way of reasoning about
propositions for which you do not have enough information to decide,
absolutely, whether each is true or false.

When you don't know if a proposition $H$ is true or false, you must
assign some \emph{degree of plausibility} to the proposition.  You
probably want that degree of plausibility to meet several desiderata,
including that \textsl{(1)}~degrees of plausibility can be represented
by real numbers, that \textsl{(2)}~they obey common-sense criteria,
like that if the plausibility of $A$ increases, the plausibility of
not-$A$ decreases and that plausibility of the joint hypothesis $A$
and $B$, for instance, be related in sensible ways to the
plausibilities of $A$ and $B$, and that \textsl{(3)}~they obey
consistency requirements, like that the plausibility depends only on
the evidence, and not the order in which that evidence is considered
or other irrelevancies.  If you place requirements like these, you are
led inexorably and provably to Bayesian inference, with posterior
probabilities taking the place of the ``degree of plausibility'' you
seek, and Bayes's Theorem relating those posterior probabilities to
likelihoods and prior probabilities.\footnote{This is demonstrated,
  among other places, in Jaynes (2003), an amusing and problematic
  text.  Jaynes gives credit for the argument to Cox.}

Because these are good desiderata, and because (as a community at
least) we are relatively rational, the progress of astrophysics really
does follow some kind of approximation to Bayesian reasoning.  We
\emph{are} Bayesians, even if we don't want to be; even those arguing
that we shouldn't be doing data analysis by Bayesian inference are,
themselves, in their scientific reasoning, Bayesian.  We assign
degrees of plausibility to hypotheses by considering our prior
knowledge (for example, the consistency of the hypothesis with other
data, other things we strongly believe to be true, and the simplicity
of the hypothesis for communication to others) and the success of the
hypothesis in explaining the observed data (this would be in the form
of something like the likelihood), and we combine the prior and the
likelihood (approximately) multiplicatively.

As an aside, if I am right, one consequence is that the Popperian view
that theories can \emph{only} be falsified is (itself)
false.\footnote{I am exaggerating Popper's much more subtle view here
  for emphasis; this should not be taken as a statement about Popper
  himself.}  In any contentious scientific issue, there are competing,
mutually exclusive hypotheses.  No observation ever completely rules
out one hypothesis, because there are finite observational
uncertainties, because any hypothesis can always be complexified in
relevant ways, and because any observation can be excused from
applicability.\footnote{If you want examples of all of these, see the
  literature of the last decade comparing detailed galaxy properties
  with the \LCDM\ model.}  Any observation that undermines (reduces
the posterior probability of) any hypothesis $H_1$ also, in doing so,
supports (increases the posterior probability of) at least some
mutually exclusive alternative $H_2$, because the probabilities of all
the conceivable mutually exclusive hypotheses must sum to unity.
Thus, contrary to what we are taught in grade school, observations do
just as much ``ruling in'' as ``ruling out''.  This realization
justifies our feeling---which Popper would not have allowed---that the
cosmic microwave background and large-scale structure observations of
this past decade strongly endorse or support the \LCDM\ model with
adiabatic initial conditions.  We are right; they do.  On the other
hand, see my comments on exhaustiveness, below, which may undermine
this argument.

As another aside, if I am right, another consequence is that we often
find ourselves working with models that are, in detail, a bad fit to
the available data.  Strictly, if we were not Bayesians, but instead
making decisions on the basis of the absolute value of the likelihood
(the probability of the data given the model) or some equivalent, we
would never continue working with models that aren't a pretty good fit
to the data.  And yet, we often perform inference with models that
are, technically, not a good fit.  This is because inference is not
possible without a model, and so we must use the best models we have,
whatever their absolute likelihoods.

\section{The impossible demands of realism}

Bayesian reasoning is the calculus of \emph{plausible} inference.
What we would all like---and I think what Blandford was getting at
with his questions---is an understanding of the fundamental processes
that govern the Universe, an understanding that is not just
\emph{useful} for calculation but an understanding that is \emph{true}
in some deeper sense.  Typically, a scientist sees the latter point as
either obvious and important, or else completely irrelevant.  I would
like to argue that we don't have a choice; there is some very clear
sense in which \emph{truth} is not what is returned by any finite
scientific investigation; all that is returned is plausibilities (some
of which become very very high), and those plausibilities relate not
directly to the truth of the hypotheses in question, but rather to
their \emph{use} or \emph{value} in describing the data.

The fundamental reason scientific investigations can't obtain literal
truth is that no scientific investigator ever has an \emph{exhaustive}
(and mutually exclusive) set of hypotheses.  Plausibility calculations
are calculations of measure in some space, which for our purposes we
can take to be the space formed by the union of every possible set of
scientific hypotheses, with their parameters and adjustments set to
every possible set of values.  Bayes's theorem tells us how to adjust
the relative probabilities of the hypotheses (and, in detail, the
relative probabilities of different parameter settings) in the set as
each new datum arrives.  This procedure is \emph{provably correct},
which is good, but it only returns the \emph{correct} hypothesis---the
hypothesis that really does generate the physical world we are
observing---when the original set of hypotheses is exhaustive.  That
is, we only get to be realists when we have considered every possible
hypothesis.  Since the hypothesis space is almost certainly infinite
in all relevant senses, realism---the belief that confidently
established scientific results are (or can be) literally and
fundamentally true statements about the world---requires infinite
computing and human resources.\footnote{For example, Kolmogorov
  (1965), among others, proved some relevant things.  Amusingly,
  Jaynes (2003) contains many absurd pieces of advice in which these
  infinities are treated as trivial.}  If you didn't perform an
infinite amount of computation, you did not find the truth (unless you
were unimaginably lucky).

This is not bad news; in fact this liberates us to take a pragmatic
view, perform finite inference with finite hypothesis spaces, and
choose the best models we have for further study.  If we go all the
way, we can even call the very best models in this sense ``true'' and
only be misleading to the most epistemologically rigorous.  A good
analogy comes from coding theory.  By far the best compression
algorithms for transmitting data losslessly over a channel involve
building generative models of the data, where sender and receiver have
both agreed on prior information, or prior probabilities for the
settings of those models.  As the message is encoded, the model is
made better and better, and the transmission of the messsage is, in
some sense, aspects of the posterior probability distribution for the
parameters of the model plus residuals.  The receiver obtains the
posterior parameters, predicts the message, and uses the residuals to
adjust it to a lossless copy.  At no point does the sender or the
receiver ever have to ask whether their model of the message is
``true''.  They only have to decide whether their model leads to a
substantial \emph{shortening} of the message.  In case you think all
of this is crazy, this is exactly how the \project{Galileo} spacecraft
data were encoded and transmitted following the failure of the
high-gain antenna.\footnote{McEliece \& Swanson (1999).}  The sender
and receiver, in this model, are not seeking the ``truth''; they are
both pragmatists.  They recognize that the exploration of a larger
model space would lead to a shorter message, but that exploring a
larger model space might violate constraints they have on
time-to-encode, time-to-decode, buffer size, or computation.
Scientists should take as their role models not priests, who reveal
truth, but signal encoders, who improve everyone's life on a daily
basis by pragmatic accomplishment.\footnote{There is a perfect analogy
  between signal encoding and Bayesian inference and the literatures
  of the two have converged.  See, for example, MacKay (2003).  That
  is, plausible inference becomes exactly identical to signal encoding
  in an odd fantasy in which the goal of a scientist is to losslessly
  communicate the observations with the shortest possible message.  If
  sender and receiver agree on prior information, that shortest
  message will consist of the posterior probabilities of models and
  model parameters, plus residuals.  For all this to work, of course,
  the protocol for communicating the posterior probabilities must be
  chosen to be one that is optimal for the mutually agreed-upon prior
  probabilities, where I am using ``optimal'' here in the sense of
  Shannon (1948).}

\section{Expansion}

It is important to note that pragmatism and plausibility are not
obstacles to extremely confident scientific conclusion.  On the
contrary, we know many things about the Universe with great certainty.
One example is that the Universe is expanding; this has been
established beyond any doubt, even when viewed in this pragmatic
light.  Consider the evidence:

Almost all galaxies are observed to have spectral shifts consistent
with redward Doppler Shifts, and more apparently distant galaxies tend
to have larger shifts.  Now with calibrated supernovae, the
distance--redshift relationship is measured with great precision and
beautifully consistent with expansion.  All measures of intensity seem
to vary with redshift consistent with the Tolman relation, as expected
if the Universe is governed by Lorentz symmetry (so the shifts can be
interpreted as Doppler Shifts).  Along the same lines, the Cosmic
Microwave Background has not just the spectrum but also the absolute
intensity of a blackbody, as expected in an expanding model.  All
observations of cosmologically distant objects are consistent with the
Universe being denser and hotter in the past.  All reasonable
cosmological solutions to general relativity involve either expansion
or contraction.\footnote{This paragraph ought to be instrumented with
  thousands of citations, but a random and unfair subsample might
  include Hubble \& Humason (1931); Mather \etal\ (1994); Songaila
  \etal\ (1994); Riess, Press, \& Kirshner (1995); Pahre, Djorgovski,
  \& de~Carvalho (1996); More, Bovy, \& Hogg (2009).}  Finally, the
only successful physical models of structure formation at present live
in an expanding background that is consistent with the observed
Doppler Shifts.

The expansion model has an extremely high likelihood---it explains
well the data---and an extremely high prior probability---it is
consistent with our knowledge from other domains such as the theories
of electromagnetism and gravitation.  Note that it has no, and has
never had any, serious alternative.\footnote{It is obligatory to
  mention ``tired light'' at this point, but only to note that it is
  not a theory advocated by any scientist; it is only a straw man
  built to illustrate the strength of the expansion hypothesis.}
Oddly that doesn't reduce our confidence.

Even if we \emph{do} find a better explanation than expansion---and I
seriously doubt that we ever could---it won't really \emph{replace}
expansion as an explanation, it will complexify, adjust, or complement
the expansion explanation.  See, for example, the replacement of
Newtonian gravity with Einsteinian gravity; Newtonian gravity is not
seen by anyone as really \emph{wrong}, it is just a quantitative limit
of the better, more complete theory.  This point plays into the
realism point as well, but I have to admit I don't know which way.  I
think it shows that scientists are not realists at heart, even when
they think they are.  But it is also related to the fact that
inference does often return good results even when models are
simplistic; this could---in some sense---be the meaning of the word
``approximation'' or ``limit''.

\section{Fractals}

\LCDM\ is a ridiculously successful physical model of the Universe on
very large scales.  It explains, simultaneously, the observations we
have that are relevant to the expansion history, the angular spectrum
of perturbations in the cosmic microwave background, the growth of
structure on large scales, and the abundances of large, collapsed
objects.\footnote{Once again, the reference list could be long here,
  but an unfair sampling of the very most recent results might include
  Tegmark \etal\ (2004), Eisenstein \etal\ (2005), and Komatsu
  \etal\ (2009).}  The large-scale structure observations and
predictions cover an enormous range in scale, cosmic time, and growth
factor.  As with expansion, it is difficult to imagine another theory
truly \emph{replacing} \LCDM: The worst-case scenario for \LCDM\ at
this point is that it will be seen forever as an exceptionally
successful approximation to the true theory; one that permits easy
calculation of a multitude of accurately observed phenomena.

There are many issues with \LCDM\ at small scales, many of which I
work on, in the hope that we will be obliged to complexify the
over-simple model and open up new space for fundamental
discoveries.\footnote{I comment on this motivation in Hogg~(2005).}
But the successes on large scales are such that any modifications to
\LCDM\ must be made carefully so as not to disturb the large-scale
successes.  In short, at this point we can see the large-scale success
of \LCDM\ as establishing good certainty for the model.

Perhaps the most trivial prediction of all for \LCDM\ (and its
physically motivated competitors, such as the DGP model discussed
below) is that the Universe has a \emph{mean density}---all physical
cosmological theories are calculated on a background that is
homogeneous on the largest scales.  In a recent set of papers, this
fundamental prediction of the model has been tested; some
investigators argue that this prediction is not consistent with the
observations.\footnote{Sylos~Labini \etal\ (2009a), Sylos~Labini,
  Vasilyev, \& Baryshev (2009b).}  The Universe does not have to have
a mean density, of course, but if it doesn't, then it is---in some
sense---a fractal, or fundamentally inhomogeneous.  My view is that
homogeneity is well tested and qualitatively and quantitatively in
agreement with \LCDM.\footnote{We performed a straightforward test in
  Hogg \etal\ (2005), designed to be insenstive to unknown issues with
  calibration and evolution.  My view is that a combination of issues
  with the data and with galaxy evolution create the results of
  Sylos~Labini \etal\ (\latin{op.\ cit.}); the Hogg \etal\ test is
  more robust.}  But let's imagine, \emph{just for the purposes of
  argument,} that the observations \emph{did} suggest, at some level,
that there is no mean density.

Fractals are beautiful and approximations to fractals (fractal-like
functions confined to a limited range of scales) are abundant in the
natural world.  Certainly the massive galaxy--galaxy auto-correlation
function is close to a single unbroken power law over all scales on
which we can measure it.\footnote{The largest range of scales is shown
  in Masjedi \etal\ (2006).}  So it is tempting to think about
inhomogeneous models.

Unfortunately---and importantly---there \emph{are no quantitative
inhomogeneous models}.  There are no solutions to general relativity
with an inhomogeneous matter distribution.  Because of inevitable
distortions to the metric and expansion history, observables are
impossible to compute (for example, even the distance--redshift
relation would become unusable), even if a background were known.  The
idea that the Universe is inhomogeneous makes \emph{no} quantitative
predictions and explains \emph{nothing}, so it is not a scientific
contender, even if the evidence against a mean density becomes quite
strong.  Note the blow this strikes against pure realism.

A fractal model can only become a contender in one of two ways: Either
the evidence against a mean density must get so strong that it
outweighs the success of every other prediction made by the
homogeneous \LCDM\ model, of which there are many.  This condition is
so far from being met, I can't see any way to meet it, even if we
measure the redshift of every luminous galaxy inside the horizon.
Alternatively, someone can devise a method to compute an inhomogeneous
model and predict a number of observables and show that the fractal
model does as well as or better than \LCDM.  I don't know enough to
know if this is possible, but I don't think there are even any
strategies for executing this ambitious theoretical program; probably
they would have to be numerical.  That said, I do not doubt that any
success in this field would have a big impact on the study of gravity
even if it doesn't turn out to be a good fit to the data.  Here is a
subject with which someone ambitious could profitably pack up and sail
on \project{The Beagle}.

Back to reality: There \emph{isn't} at present good evidence against a
mean density; homogeneity is well established and in agreement with
the \LCDM\ predictions.  An inhomogeneous universe is so intractable
that there is almost no near-term future in which we are likely to be
able to either observe or compute anything interesting in this area.
This is an apparent controversy, but in fact great confidence is
warranted.  The potentially disturbing aspect of the story I have told
here is that the confidence comes in part from the intractability
itself!  But of course a pragmatist is perfectly happy with that.

\section{Large extra dimensions}

Despite the great success\footnote{Actually, some of my colleagues
  would say that \LCDM\ is a \emph{failure} because it makes use of a
  cosmological constant that it so far from either a particle-inspired
  value or zero that it is extremely implausible \latin{a priori}.
  That is somehow related to all of this.} of the standard
\LCDM\ model---or perhaps because of it---the world of fundamental
cosmology is bristling with new ideas.  One of the most interesting
new ideas is that 3+1-dimensional gravity is just what we observe of
some higher-dimensional model because we are (somehow) confined to a
lower-dimensional subspace.  The most worked-out example of this is
the DGP model, which contains a simple idea but for which it has been
challenging to do precise calculations.\footnote{The DGP model was
  introduced by Dvali, Gabadadze, \& Porrati (2000); a cosmological
  (homogeneous) solution was found by Deffayet (2001); and growth of
  structure was calculated by Hu \& Sawicki (2007) and Scoccimarro
  (2009) among others.}  Because the current cosmological solution in
DGP has a single parameter that determines the properties of the
background expansion, it has the same global freedom as the
\LCDM\ model; that is, the two models have (more-or-less) the same
number of parameters.  In a straightforward comparison between GDP and
\LCDM\ and a basket of observations there is $\Delta\chi^2\equiv
2\,\Delta\ln\like\approx 20$, in favor of \LCDM.\footnote{This
  likelihood ratio comes from Fang \etal\ (2008); a very different
  conclusion is reached by Sollerman \etal\ (2009) with a more careful
  analysis but of a smaller data set.}  On the face of it, without
evaluating the calculation of the likelihood ratio or any controversy
related thereto, this is very bad news for DGP.  Is the existence of
large extra dimensions ruled out at high confidence?

There are considerations here that prevent any trivial answer to this
question.  The first is that neither model is a good fit to the data.
This means either \textsl{(1)}~that both models would lose a model
comparison with some better, third model, or \textsl{(2)}~that both
models must be complexified fundamentally with additional physics, or
\textsl{(3)}~that the observational uncertainties have been
under-estimated.  In the first case, it is irrelevant that \LCDM\ is
preferred to DGP, because \LCDM\ is disfavored over all; that is, the
test between \LCDM\ and DGP does not establish \LCDM\ against its
truer competitor.  In the second case, the addition of new physics
will inevitably give both models more freedom; it is not clear which
model will more naturally obtain the freedom necessary to obtain a
substantially higher likelihood when compared with the observations.
The model comparison, at present, is between two models that have had
certain relevant physics ``switched off'' and the amount that affects
each model is likely to be different.  In the third case, the
investigator is encouraged to think generally about what can be wrong
with the observational uncertainties: Are there likely to be some data
points that are rejectable outliers?  Do all data points have
underestimated uncertainties?  Which data points are qualitatively
similar or similar in origin?  Once these questions are answered, the
uncertainties \emph{in} the observational uncertainties must be
modeled, with parameters fit simultaneously with the fundamental
cosmology parameters, and then marginalized out for the model
comparison.  This procedure, if performed symmetrically for the two
models (as it must be, since it relates to the data alone) will
inevitably reduce the magnitude of the relative likelihood of the two
models.

The second complexity relates to the issue that the DGP model as
currently calculated is a very specific and highly non-linear theory,
not all the details of which are understood.  The calculation of the
growth of structure is probably much more general than DGP, in the
sense that it is a calculation in an effective theory generated by
DGP, that could in principle be generated by many other fundamental
theories.  But it is also the case that the DGP theory, in a different
background or with different brane properties, might be able to
support other effective theories for the growth of structure.  That
is, a test of the growth of structure in the DGP model as currently
calculated is not a direct test of the existence of large extra
dimensions itself; it is a test of an extra-dimensions-motivated
alternative gravitational theory.  There is a lot of theory to be done
to bring us closer to understanding what fundamental modifications to
gravity (subject to Solar System and other constraints) bring about
what effective cosmological theories, and what the predictions are for
each of those effective cosmological theories.  Unfortunately, given
the precision of contemporary data, only a small family of effective
theories is going to end up being consistent with the data; much of
this time-consuming theoretical work will lead to important but null
results.

\section{The multi-armed bandit}

These considerations play into the decision-making of what Blandford
called the ``young fogeys'': Although putting a
sustained effort into calculating and testing alternative gravity
theories could result in some beautiful physics, until it looks like
it has a good chance of raising the posterior probability of some
model above the orthodoxy, that career segment may languish uncited
and lacking in influence on the business.  The ``old turks'' have more
to spend and less to lose.

There is a class of problems in decision-making known as ``multi-armed
bandits'' in which a gambler is presented with a machine with $K$
levers, analogous to a slot machine with one lever, where each of the
$K$ levers has different and unknown probabilities for different
payoffs.\footnote{This problem was effectively introduced by Robbins
  (1952); there is now a long literature on strategies.}  At each
round, the player must decide which lever to pull, where each pull may
provide some reward and will definitely provide some information.  The
details of a player's strategy can depend strongly on the player's
\emph{utility}, or what the player is trying to achieve.  This will
depend, in turn, on things like tolerance for risk, cost of lever
pulling, and discount rate for future cash flow.

This toy problem---as odd as it sounds---is an analogy for the
performing of scientific investigations, and indeed the problem was
initially raised in the context of adaptive experimental design.
Every morning, an investigator must decide what project to concentrate
on---what ``lever to pull''---whether to work on incremental
improvements to an orthodox model or develop or test some radical
model; the investigator must make this decision without knowing
precisely how much he or she will be paid (or charged) for the choice.
And, perhaps disturbingly (but related to issues of exhaustiveness
above), the number of levers that the investigator can pull is far,
far larger than the number of times he or she gets to pull one: There
are far more good ideas (at least in cosmology) than there is
investigator-time to explore them.\footnote{And, of course, the levers
  change their payoff distributions as context changes, and new levers
  become available as new scientific opportunities arise, so the pure
  multi-armed bandit problem is actually far less general than the
  dilemmas of a scientist.}  The investigator's utility is the key.
In work on experimental design, the utility is usually imagined to be
purely related to information or knowledge about the question at hand.
But in real decision-making, the utility involves not just questions
of knowledge, but also of real-world costs and benefits.  In these
matters, the utilities are very different for young and old
scientists, where younger scientists ought to have less tolerance for
risk and a higher discount rate for future payoffs (because they need
to get their PhD or postdoc or faculty job or tenure with a short time
horizon), and older scientists ought to have more tolerance and a
lower discount rate.  Hence the relative orthodoxy of the youth.

Despite all this, at present there are in fact a substantial number of
the youth working on ideas as speculative as extra dimensions and
fractals and even more speculative.  Some of my young colleagues are
looking for observational signatures of universe--universe collisions!
It is possible that our community has recovered somewhat from
Blandford's complaint.  There are many respects in which cosmology
appears far more exciting now than it did in the nineteen-nineties,
despite the fact that the \LCDM\ parameters have all been constrained
so well that cosmology is now referred to as a ``mature science''.  I,
for one, have learned that there is life even in middle age.

\paragraph{Acknowledgements:}
Much of the material presented here was informed by conversations over
many years with John Bahcall, Roger Blandford, Jo Bovy, Dustin Lang,
Phil Marshall, Jim Peebles, Hans-Walter Rix, and Sam Roweis.  I got
specific help on this manuscript from Jo Bovy, Gregory Gabadadze, Lam
Hui, Dustin Lang, and Roman Scoccimarro.  Support was provided by NASA
(grant NNX08AJ48G), the NSF (grant AST-0908357), and a research
fellowship from the Alexander von~Humboldt Foundation.

\section*{References}
\begin{list}{}{%
  \setlength{\leftmargin}{\parindent}
  \setlength{\rightmargin}{0em}
  \setlength{\itemindent}{-\parindent}
  \setlength{\parsep}{0.5ex}
  \setlength{\itemsep}{0ex}
}\footnotesize
\item
  Deffayet,~C., 2001,
  Cosmology on a brane in Minkowski bulk,
  \textit{Physics Letters B} \textbf{502} 199--208.
\item
  Dvali,~G.~R., Gabadadze,~G., \& Porrati,~M., 2000,
  4D gravity on a brane in 5D Minkowski space,
  \textit{Physics Letters B} \textbf{485} 208--214.
\item
  Eisenstein,~D.~J. \etal, 2005,
  Detection of the baryon acoustic peak in the large-scale correlation
  function of \project{Sloan Digital Sky Survey} Luminous Red Galaxies,
  \textit{The Astrophysical Journal} \textbf{633} 560--574.
\item
  Fang,~W., Wang,~S., Hu,~W., Haiman,~Z., Hui,~L., \& May, M. 2008,
  Challenges to the DGP model from horizon-scale growth and geometry,
  \textit{Physical Review D} \textbf{78} 103509.
\item
  Hogg,~D.~W., Eisenstein,~D.~J., Blanton,~M.~R., Bahcall,~N.~A.,
  Brinkmann,~J., Gunn,~J.~E., \& Schneider,~D.~P., 2005,
  Cosmic homogeneity demonstrated with luminous red galaxies,
  \textit{The Astrophysical Journal} \textbf{624} 54--58.
\item
  Hogg,~D.~W., 2005,
  What best constrains galaxy evolution in the local Universe?,
  arXiv:astro-ph/0512029.
\item
  Hu,~W. \& Sawicki,~I., 2007,
  Parametrized post-Friedmann framework for modified gravity,
  \textit{Physical Review D} \textbf{76} 104043.
\item
  Hubble,~E. \& Humason,~M.~L., 1931,
  The velocity-distance relation among extra-galactic nebulae,
  \textit{The Astrophysical Journal} \textbf{74} 43--80.
\item
  Jaynes,~E.~T., 2003,
  \textit{Probability Theory:\ The Logic of Science}
  (Cambridge University Press).
\item
  Kolmogorov,~A.~N., 1965,
  Three approaches to the quantitative definition of information,
  \textit{Problems in Information Transmission} \textbf{1} 1--7.
\item
  Komatsu,~E. \etal, 2009,
  Five-year \project{Wilkinson Microwave Anisotropy Probe}
  observations:\ Cosmological interpretation,
  \textit{The Astrophysical Journal Supplement} \textbf{180} 330--376.
\item
  Mackay,~D.~J.~C., 2003,
  \textit{Information Theory, Inference, and Learning Algorithms}
  (Cambridge University Press).
\item
  Masjedi,~M. \etal, 2006,
  Very small-scale clustering and merger rate of luminous red galaxies,
  \textit{The Astrophysical Journal} \textbf{644} 54--60.
\item
  Mather,~J.~C. \etal, 1994,
  Measurement of the cosmic microwave background spectrum by the
  \project{COBE FIRAS} instrument,
  \textit{The Astrophysical Journal} \textbf{420} 439--444.
\item
  McEliece,~R.~J. \& Swanson,~L., 1999,
  Reed--Solomon Codes and the exploration of the Solar System,
  in \textit{Reed--Solomon Codes and Their Applications},
  Wicker,~S.~B. \& Bhargava,~V.~K., eds. (Wiley-IEEE Press) 25--40.
\item
  More,~S., Bovy,~J., \& Hogg,~D.~W., 2009,
  Cosmic transparency:\ A test with the baryon acoustic feature
  and type Ia supernovae,
  \textit{The Astrophysical Journal} \textbf{696} 1727--1732.
\item
  Pahre,~M.~A., Djorgovski,~S.~G., \& de~Carvalho,~R.~R., 1996,
  A Tolman surface brightness test for universal expansion and the
  evolution of elliptical galaxies in distant clusters,
  \textit{The Astrophysical Journal Letters} \textbf{456} L79--L82.
\item
  Riess,~A.~G., Press,~W.~H., \& Kirshner,~R.~P., 1995,
  Using Type IA supernova light curve shapes to measure the Hubble constant,
  \textit{The Astrophysical Journal Letters} \textbf{438} L17--L20.
\item
  Robbins,~H., 1952,
  Some aspects of the sequential design of experiments,
  \textit{Bulletin of the American Mathematical Society} \textbf{58} 527--535.
\item
  Scoccimarro,~R., 2009,
  Large-scale structure in brane-induced gravity I.\ Perturbation theory,
  arXiv:0906.4545.
\item
  Shannon,~C.~E., 1948,
  A mathematical theory of communication,
  \textit{Bell System Technical Journal} \textbf{27} 379--423, 623--656.
\item
  Sheldon,~E.~S. \etal, 2009,
  Cross-correlation weak lensing of \project{SDSS}
  galaxy clusters.~I.~Measurements,
  \textit{The Astrophysical Journal} \textbf{703} 2217--2231.
\item
  Sollerman,~J. \etal, 2009,
  First-year \project{Sloan Digital Sky Survey-II} supernova
  results:\ Constraints on nonstandard cosmological models,
  \textit{The Astrophysical Journal} \textbf{703} 1374--1385.
\item
  Songaila,~A. \etal, 1994,
  Measurement of the microwave background temperature at a redshift of 1.776,
  \textit{Nature} \textbf{371} 43--45.
\item
  Sylos~Labini,~F., Vasilyev,~N.~L., \& Baryshev,~Y.~V., 2009b,
  Breaking the self-averaging properties of spatial galaxy fluctuations in
  the \project{Sloan Digital Sky Survey} Data Release Six,
  arXiv:0909.0132.
\item
  Sylos~Labini,~F., Vasilyev,~N.~L., Baryshev,~Y.~V.,
  \& Lopez-Corredoira,~M., 2009a,
  Absence of anti-correlations and of baryon acoustic oscillations in
  the galaxy correlation function from the
  \project{Sloan Digital Sky Survey} DR7,
  arXiv:0903.0950.
\item
  Tegmark,~M., \etal, 2004,
  Cosmological parameters from \project{SDSS} and \project{WMAP},
  \textit{Physical Review D} \textbf{69} 103501.
\end{list}

\end{document}